\documentclass[prb,twocolumn,showpacs,amsmath,amssymb,floatfix]{revtex4}
\usepackage{graphicx}
\usepackage{ulem}
\usepackage{bm}
\usepackage{amsmath}
\usepackage{latexsym}
\usepackage{amsfonts}
\usepackage{amssymb}
\usepackage{verbatim}
\usepackage{color}
\usepackage{verbatim}
\makeatletter

\newcommand{\sign}{\text{sgn}}
\DeclareMathOperator{\tr}{Tr}
\allowdisplaybreaks

\makeatother
\begin{document}
\title{Time-resolved spectroscopy at surfaces and adsorbate dynamics:\\ 
insights from a model-system approach\\}
\date{\today}
\author{Emil Bostr\"om$^1$, Anders Mikkelsen$^2$, and Claudio Verdozzi$^1$}
\affiliation{$^1$Mathematical Physics and ETSF, Lund University, Box 118, S-22100 Lund, Sweden\\
$^2$ Synchrotron Radiation Research, Lund University, Box 118, S-22100 Lund, Sweden}

\begin{abstract}
We introduce a model description of femtosecond laser induced desorption at surfaces. The substrate part of the system 
is taken into account as a (possibly semi-infinite) linear chain. Here, being especially interested to the early stages
of dissociation, we consider a finite-size implementation of the model (i.e. a finite substrate), for which
an exact numerical solution is possible. By time-evolving the many-body wavefunction, and also using results from a novel time-dependent DFT description for electron-nuclear systems, 
we analyse the competition between several surface-response mechanisms and electronic correlations in the transient and longer time dynamics under the influence of dipole-coupled fields. 
Our model allows us to explore how coherent multiple-pulse protocols can impact desorption in a variety of prototypical experiments.
\end{abstract}

\maketitle

\section{Introduction}
Femtosecond (fs) laser technology has revolutionised our understanding and control of chemical reactions \cite{Zewail}.
However, it also presents serious theoretical challenges, since ultrafast measurements probe 
atomic/molecular scales far away from equilibrium, where an accurate description of the concerted motion of electrons and nuclei is indispensable to interpret the experiment. Progress has been made for free molecules \cite{FarisG,Henriksen,Gisselbrecht,Remacle,PerStef}, but a comparable understanding is still lacking for surfaces \cite{Frischkorn,Petek1}. This is unfortunate, since many important reactions are catalysed by a surface \cite{Gabor,Hornett,Linic}, e.g. photocatalytic processes directly involving light-matter interaction \cite{SurfScience}.
A key technology for ultrafast studies are Ti:Sapphire lasers \cite{Frischkorn,Keller,Sutter}: With a central frequency of 800nm and pulse durations from hundreds to a few fs, these lasers have been pivotal to the emergence of novel surface-sensitive ultrafast microscopies/spectroscopies \cite{Aeschlimann,Marsell} and to new results on chemical reactions and desorption. \cite{Bartels,Backus,KiHyun,Muino,Fuchsel,Petek2,Nuernberger,Tremblay,DellAngela}. 

An accurate first principle description of ultrafast dynamics at surfaces is within the scope of comprehensive approaches such as
the non-equilibrium Green's function (NEGF) method \cite{BalzerBonitzbook,bookNEG} and Time-Dependent Density Functional Theory (TDDFT) \cite{RG84,EsaRas,Hellgren}, where there is ongoing effort in this direction \cite{Gross,RvL2,Butriy,Tavernelli,Prezhdo,dipole,Stefanucci,Perfetto1}. However, with current treatments of electron-electron and electron-nuclear interactions, the inherently non-perturbative situation of desorption is in general not adequately described even in the initial stages.
Thus, it remains highly relevant to theoretically explore simplified models which simulate experiments with fs lasers, and utilize the possibilities of control offered by their pulse structure \cite{KiHyun,Muino}. 

Motivated by this, we present here a model approach to pump-probe real-time dynamics of adsorbates, incorporating electron interactions, core-hole relaxation, plasmon screening and anharmonic nuclear dynamics. Our investigation focusses on the early stages of dissociation dynamics; however, in the rest of the paper, this specific sub-regime we be simply referred to as "desorption".
Our novel description merges elements from three popular surface-physics models: the Anderson-Newns-Grimley model of chemisorption \cite{Anderson,Newns,Grimley}, the charge-transfer model of core photoemission \cite{Langreth70,KotaToyo,Gumhalter,SchonGun,CiniSS79,Sugano,Brenig} and the Shin-Metiu model of nuclear motion \cite{ShinMetiu}. By considering finite systems, we exactly and simultaneously address several competing time-scales and response mechanisms, to gain robust, albeit qualitative, insight in the ultrafast regime.

Our main results are:
i) For short pulses, it appears a unified treatment of 
electrons and (light) nuclei is needed already for the early stage of the dynamics.
ii) Desorption can be controlled in experimentally viable pulse protocols by manipulating the plasmon response.
iii) A multicomponent TDDFT description of adsorbate dynamics unveils highly non-trivial features in the Kohn-Sham potentials.
Overall, the results show that our model provides insight into a wide range of situations for adsorbate dynamics (accessible by measuring e.g. the adsorbate-surface bond length or the change in desorption yield), and 
represents a novel and versatile benchmark for more realistic theoretical treatments.

The paper is structured as follows: In Section~\ref{sec:model} we define the Hamiltonian of our model, and discuss its parameters and the limitations of its finite size realization. In Section~\ref{sec:equilibrium} we present results pertaining to the equilibrium properties of the system, such as energy level structure and spectral and response functions. In Section~\ref{sec:desorption} we discuss general dynamical properties of the system, and show how different pulse protocols can be utilized to control the evolution of the adsorbate. To get insight into the results we in Section~\ref{sec:tddft} present a novel version of TDDFT and show the exact Kohn-Sham potentials for both the electrons and the nuclei.

%
%%%%%%%%%%%%%%%%%%%%%%%%%%%%%%%%%%%%%%%%%%%%%%%%%%%
%%%%%%%%%%%%%%%%%%%%%%%%%%%%%%%%%%%%%%%%%%%%%%%%%%%
%%%%%%%%%%%%%%%%%%%%%%%%%%%%%%%%%%%%%%%%%%%%%%%%%%%
%%%%%%%%%%%%%%%%%%%%%%%%%%%%%%%%%%%%%%%%%%%%%%%%%%%
%%%%%%%%%%%%%%%%%%%%%%%%%%%%%%%%%%%%%%%%%%%%%%%%%%%
%%%%%%%%%%%%%%%%%%%%%%%%%%%%%%%%%%%%%%%%%%%%%%%%%%%
%

\begin{figure}[t]
\includegraphics[width=0.47\textwidth]{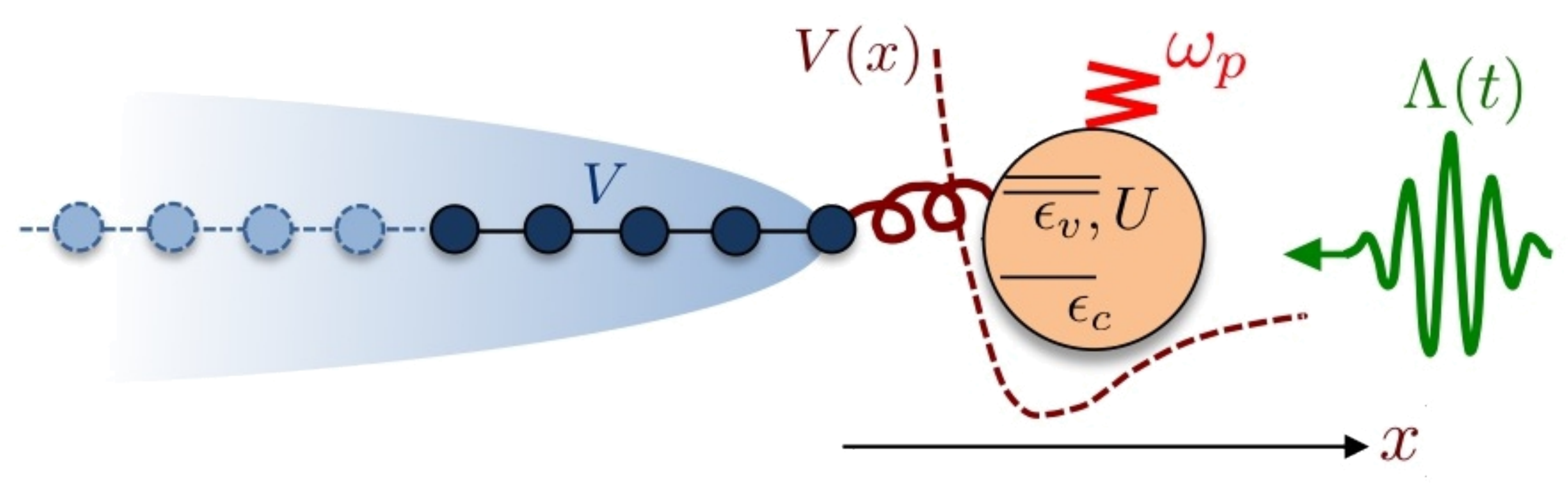}
\caption{The model of Eq.~(\ref{Ham0}), for a mobile adsorbate 
(with one core and two interacting valence levels) on a 5-site substrate. 
The adsorbate charge fluctuations are coupled to a local/surface plasmon. 
The substrate-adsorbate effective potential and the laser perturbation are also
schematically shown.}
\label{fig:model}
\end{figure}

\section{The model}\label{sec:model}
The system we consider is illustrated in Fig.~\ref{fig:model}, and consists of a rigid chain (the substrate) of $L$ sites with one orbital per site and with a mobile adsorbate at one end. The Hamiltonian is given by
\begin{align}\label{Ham0}
\hat{H}(t) = \hat{H}_s + \hat{H}_a + \hat{H}_{as} + \hat{\Lambda}(t),
\end{align}
with $\hat{H}_s$ describing the substrate, $\hat{H}_a$ the adsorbate, $\hat{H}_{as}$ the adsorbate-substrate interaction, and where $\hat{\Lambda}(t)$ is the external laser field.
To substrate Hamiltonian is taken to be
\begin{align}\label{Ham_s}
\hat{H}&_s = - V\!\!\!\sum_{\langle RR'\rangle,\sigma} c_{R,\sigma}^\dagger c_{R',\sigma}+\omega_p b^\dagger b
\end{align}
where $c_{R,\sigma}^\dagger$ creates an electron with spin $\sigma$ at site $R$ of the substrate (we use the index $S$ for the ``surface'' site of the substrate), $V$ is the hopping amplitude in the chain and $b^\dagger$ is the creation operator of a plasmon with frequency $\omega_p$. We consider only nearest neighbor hopping, which is indicated by the braces around the summation indexes.

The adsorbate Hamiltonian $\hat{H}_a$ is given by
\begin{align}\label{Ham_a}
\hat{H}&_{a} = \epsilon_c \hat{n}_c + \frac{\hat{p}^2}{2M} + \sum_{v,\sigma} \epsilon_v \hat{n}_{v,\sigma}, \nonumber \\
& + \sum_{vv',\sigma\sigma'} U_{vv'} \hat{n}_{v,\sigma} \hat{n}_{v',\sigma'} -w (1-\hat{n}_c) \hat{N}_{a}
\end{align}
where $\hat{x}$ and $\hat{p}$ denote the position and momentum operators for an adsorbate with mass $M$, and the operator $a_{v, \sigma}^\dagger$ creates an electron with spin $\sigma$ and energy $\epsilon_v$ in the valence orbital $v$ of the adsorbate. We denote by $\hat{n}_{v,\sigma} = a_{v, \sigma}^\dagger a_{v, \sigma}$, $\hat{n}_{R,\sigma}= c_{R \sigma}^\dagger c_{R, \sigma}$ and $\hat{n}_c$ single level number operators, the latter for a structureless core level of energy $\epsilon_c$ on the adsorbate, and introduce the total-number operator $\hat{N}_a=\hat{n}_c+\sum_{v,\sigma} \hat{n}_{v,\sigma} $ of the adsorbate and its ground state value $\langle \hat{N}_a \rangle_0$. The adsorbate has two valence levels (with one exception discussed in relation to Fig.~\ref{fig:TDDFT}) where electrons interact mutually with strength $U_{vv'}$ (with $U_{11}=2 U_{12} =U_{22}=U$). In case of core-hole photoemission or Auger recombination, the valence electrons experience an additional interaction $w$ (acting as a local potential), which depends on the core-level occupation \cite{SchonGun}. Since the system is always in a state with 0 or 1 core electrons \cite{Langreth70} we have $\hat{H}\equiv \hat{H}(n_c)$, and we consider the case of $N_e = L +1$ spin-compensated electrons in the other orbitals.

The adsorbate-substrate interaction Hamiltonian $\hat{H}_{as}$ is given by 
\begin{align}
\hat{H}&_{as} = \frac{\kappa}{\hat{x}^4} 
- g  e^{-\lambda (\hat{x}-1)} \sum_{v,\sigma}\left(a_{v,\sigma}^\dagger c_{S,\sigma} + h.c.\right) \nonumber \\
&+ \gamma \left(\hat{N}_{a} -\langle \hat{N}_{a}\rangle_0\right)\left(b^\dagger+b\right) \label{Ham_as}
\end{align}
where the first term gives a repulsive ion-ion interaction, and the second term is attractive \cite{Pettifor} and due to electron hopping between the adsorbate and surface sites, whose probability decays exponentially with distance \cite{surfacebook}. Together they create a Morse-like potential landscape for the adsorbate, and we consider the parameters $\kappa$, $\lambda$ and $g$ as phenomenological to give a reasonable binding energy $E_b$, vibrational frequency $\omega_{ph}$ and effective hopping amplitude $V_{e} = \langle ge^{-\lambda (\hat{x}-1)}\rangle$ \cite{surfacebook}.

The effective plasmon response to charge fluctuations on the adsorbate is given by the last term in $H_{as}$, where the value of $\langle \hat{N}_a \rangle_0$ is to be found self-consistently and $\gamma$ determines the electron-plasmon coupling strength.

The external field $\hat{\Lambda}(t)$ depends on the experiment considered, but is restricted by
$\hat{\Lambda}=0$ for $t\le0$. In the dipole approximation 
\begin{align}
\hat{\Lambda}(t) = \sum_{v \neq v',\sigma} \Lambda_{vv'}(t) a_{v\sigma}^\dagger a_{v'\sigma},
\end{align}
and induces transitions between the valence levels. Core-hole photoemission is treated in the sudden limit (see e.g. \cite{Almbladh85,Pavlyuk15}), where $\hat{\Lambda}(t)$ moves the core occupation from 1 to 0 at time $\tau_c$, to mimic the promotion of a core electron to a continuum state (left out of the explicit description).

Parameters used in the model can be obtained for specific systems using first principle simulations or experimental observations. As a result, a possible use of our model is to address qualitative features in specific realistic systems and experimental conditions. Here, however, we are interested in demonstrating the generic usefulness of the model for addressing the early stages of surface-adsorbate dynamics, and have therefore chosen typical physical parameter values, as discussed in the following subsection.

\subsection{Parameters}\label{sec:parameter}
To set our energy scale we in the following take $V = 1$, whose value is typically $V \simeq 1-2$ eV in a metal \cite{Boudeville,Taneda}. The plasmon frequency can vary substantially, but typical values as measured experimentally are in the range $4-10$ eV for surface modes \cite{Pitarke}, consistent with the classical results $\omega_p^2 = 4\pi ne^2/m$ and $\omega_s = \omega_p/\sqrt{2}$ for bulk and surface modes respectively. In the following we take $\omega_p = 4$ corresponding to a typical frequency.

Typical values of the diagonal interaction integrals $U_{vv} = U$ are in the range $1-10$ eV for adsorbed atoms \cite{surfacebook,Hubbard,Anderson}, with the off-diagonal elements slightly smaller. Increasing the interaction strength enhances the degree of correlation among the electrons, and a common indicator of strong correlations is when the ratio $U/W$ becomes larger than unity, with $W$ the bandwidth of the system (for the one-dimensional Hubbard model $W=4t$). In this paper we consider $U=1$ and $U=4$, where the latter gives $U/W = 1$, i.e it marks the onset of the strong correlation regime. 

The energy levels $\epsilon_v$ could in principle be taken both below, both above or one below and one above the Fermi level of the chain, depending on the system of interest (corresponding to anionic, cationic or neutral adsorption respectively \cite{Lang}, assuming they would be filled for an isolated atom). Here we take as a possible (and plausible) value $\epsilon_v = -U/2$ for the lower level and adjust $\epsilon_v$ of the upper level to obtain half-filling on the adsorbate in the ground state. 
The core level is assumed to be deep, and its role comes from the Coulomb stabilization energy $w$ that acts as a local potential on the valence levels, and which is typically $5-10$ eV in magnitude \cite{Ueba}. Here we have taken $w=6$ in order for core level emission to have substantial impact, as is the case in many naturally occurring situations. The role of the mass $M$ is to set the time-scale of the nuclear dynamics, and should in principle be adjusted depending on the species of atoms considered. Since we focus here on qualitative features we have chosen the rather small value $M = 352$, roughly corresponds to the mass of Hydrogen, in order to have reasonable simulation times and in line with earlier approaches \cite{ShinMetiu}.

Our simulations were performed with the bare values $\kappa' = 0.3$, $g' = 6$ and $\lambda' = 2$, after which the interatomic coordinate is rescaled in order to measure length in units of the equilibrium distance $x_{eq}$. For a chain with $L = 5$ and $U = 4$, this is equivalent to using the values $\kappa = \kappa'/x_{eq}^4 = 2.42$, $g = g'e^{-\lambda' x_{eq}} = 1.83$ and $\lambda = \lambda' x_{eq} = 1.19$, the correspond to a binding energy $E_b \simeq 1.5$ (or equivalently $E_b \simeq 1.5 -3$ eV), a phonon frequency $\omega_{ph} \simeq 0.24$ and an effective adsorbate-surface hopping $V_{e} \simeq 1.8$, that are values typical for chemisorption in the surface molecule limit.  We take $\gamma = 1$ to have an image potential shift of $v = \gamma^2/\omega_p = 0.25$, corresponding to intermediate coupling regime.

%%%%%%%%%%%

\begin{figure}[htb]
        \includegraphics[scale=0.95]{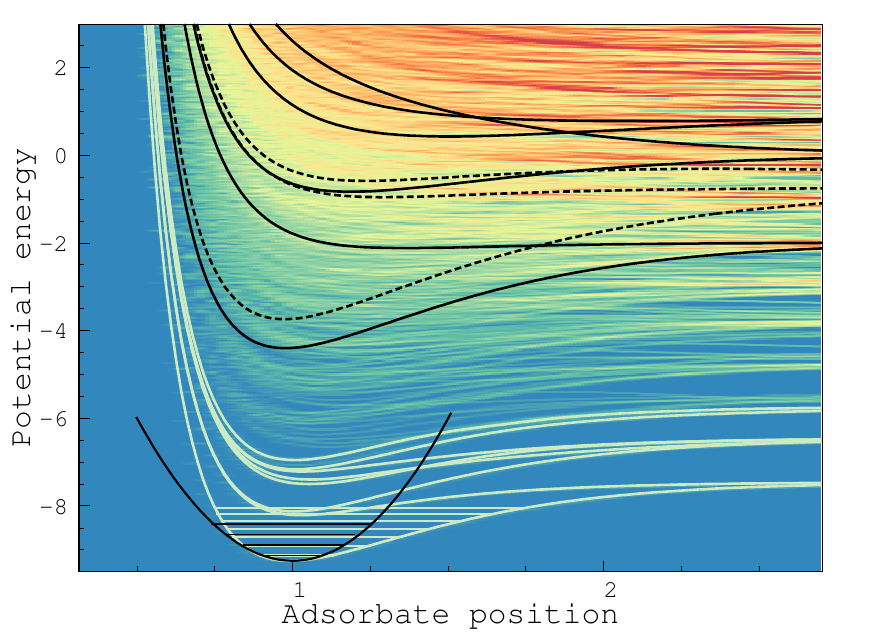}
\caption{Adiabatic potential energy surfaces (PES) for the model system of Eq.~(\ref{Ham0}) with one core and two valence levels on the adsorbate. 
The heat map is for a system with $L=5$, $U=4$, $\epsilon_1=-2$ and $\epsilon_2=1$, with colours related to the density of PES ranging from high (red) to low (blue). The light green curves show the eight lowest PES, with nuclear energy levels for the ground state potential surface and a parabolic fit (black curve) superimposed.  For comparison, the lowest PES for $L=1$ are also shown, for adsorbate interaction $U=1$ (black dashed curves) and $U=4$ (black solid curves). For $U=1$, the valence levels are at $\epsilon_1=-0.75$ and $\epsilon_2=2$ to provide adsorbate level fillings similar to $U=4$.}
\label{fig:pes}
\end{figure}

\subsection{Limitations of the model} 
Our model is subject to some limitations, the most obvious one being i) the finite size of the substrate, containing only a limited number of de-excitation channels (this however, besides making possible
an exact solution, can have direct relevance for dynamics on thin films \cite{howeverthinfilm}). 
Additionally, 
ii) Auger recombination \cite{Weightman,CVAuger,Moretti} is not considered. 
iii) Surface-adsorbate hopping induced by a surface plasmon should also be included (see e.g. \cite{CiniSS79}), especially for ionic chemisorption.
iv) The plasmon coupling $\gamma$ should depend on the adsorbate distance $x$, introducing an electron-nuclear-plasmon coupling term in $\hat{H}$.
v) The electronic interactions are kept local, but longer range interactions between substrate and adsorbate can in fact play an important role.
vi) Lattice vibrations and electronic interactions in the substrate are not included.
Avoiding ii-vi) corresponds to easy but computationally demanding extensions and is deferred to future work. On the other hand, to avoid i), a semi-infinite substrate can be taken into account via e.g. NEGF or TDDFT \cite{BostromPNGF}. However, in this case approximate treatments of interactions usually need to be introduced. The finite-size version of our model then provides a natural exact benchmark to such treatments.

Finally, typical of real-time dynamics approaches (and even for semi-infinite substrates) with 
finite-timespan simulations it is not possible to fully exclude that
at long times an atom re-adsorbs onto the surface after substantial bond stretching.
However, the displacement of the adsorbate in the sub-picosecond regime (as investigated here) is a prerequisite for complete desorption at later times, and the qualitative trends seen in this phase should also be reflected in desorption measurements \cite{Frischkorn,Petek1,Nuernberger}.

%
%%%%%%%%%%%%%%%%%%%%%%%%%%%%%%%%%%%%%%%%%%%%%%%%%%%
%%%%%%%%%%%%%%%%%%%%%%%%%%%%%%%%%%%%%%%%%%%%%%%%%%%
%%%%%%%%%%%%%%%%%%%%%%%%%%%%%%%%%%%%%%%%%%%%%%%%%%%
%%%%%%%%%%%%%%%%%%%%%%%%%%%%%%%%%%%%%%%%%%%%%%%%%%%
%%%%%%%%%%%%%%%%%%%%%%%%%%%%%%%%%%%%%%%%%%%%%%%%%%%
%%%%%%%%%%%%%%%%%%%%%%%%%%%%%%%%%%%%%%%%%%%%%%%%%%%
%

\begin{figure}[htb]
        \includegraphics[scale=0.55]{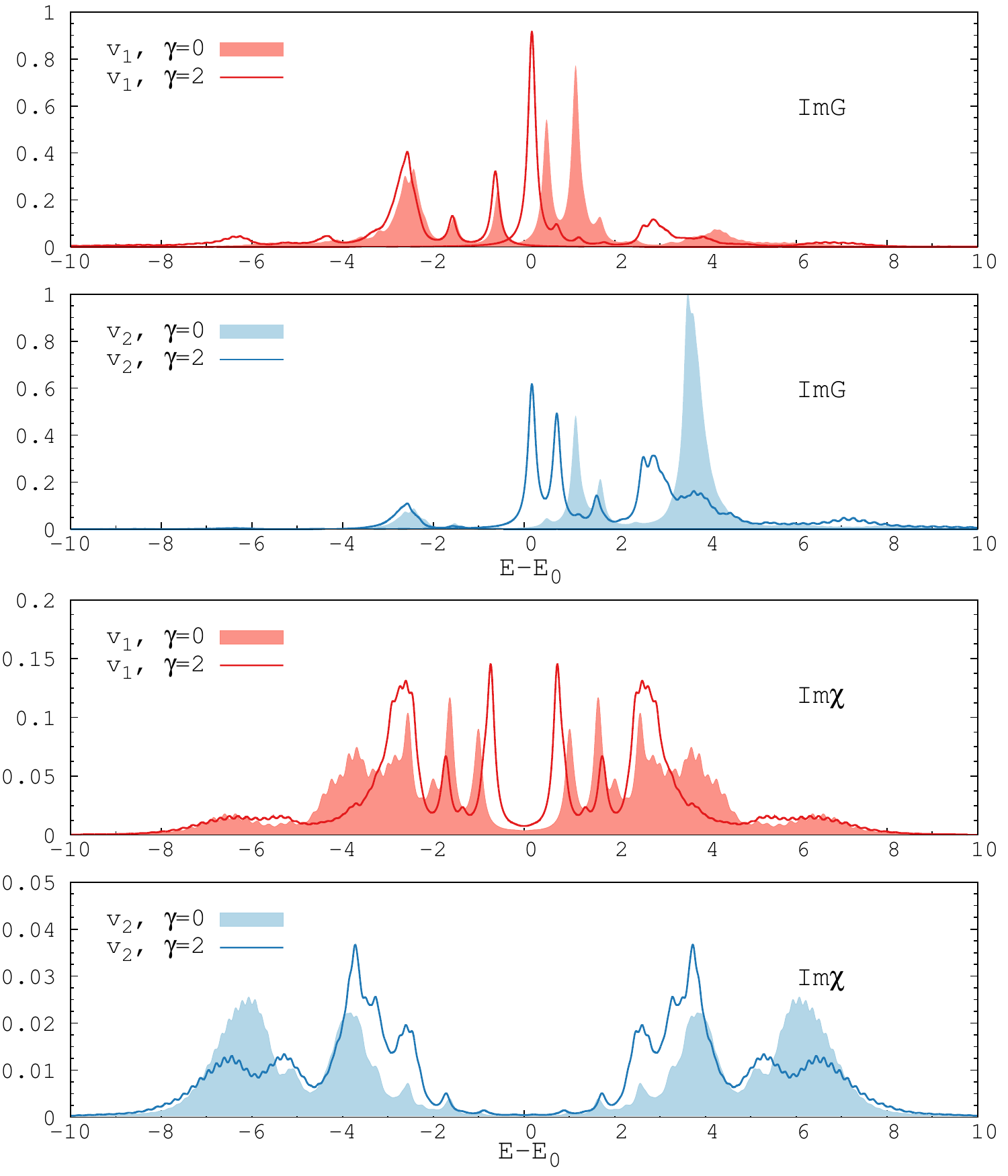}
\caption{The imaginary part of the local Green's function $G_{vv}$ and density-density response function $\chi_{vv}$ is shown for the two valence levels $v_1$ (lower) and $v_2$ (upper). All panels show the same quantity without (shaded regions) and with (lines) a local plasmon mode of frequency $\omega_p = 4$ and coupling $\gamma = 1$.}
\label{fig:spectral}
\end{figure}

\section{The equilibrium case}\label{sec:equilibrium}
In this section we discuss some equilibrium properties of the system, for the parameters $L = 1$ and 5 and $U = 1$ and 4. To find the exact ground (and initial) state $|g \rangle$ we diagonalise $H(0)$ in the basis $\{ | n_{i\sigma}, x_k, n_b \rangle \}$, where the $n_{i\sigma}$:s are site/orbital/spin occupations, $x_k$ denotes the $k$-th mesh point on a uniform grid, and $n_b$ is the plasmon occupation number.

\subsection{Potential energy surfaces}
The heat map in Fig.~\ref{fig:pes} represents the density of potential energy surfaces (PES) for $L=5$, obtained via binning around each adsorbate position (for each $x_k$,
there are 3775 PES). We see a large number of surface crossings starting already at low energies, that begin to merge at larger energy to form a quasi-continuum. The black curve shows the harmonic approximation to the ground state PES (corresponding to a phonon energy $\omega_{ph}=0.24$) that however breaks down almost immediately, as it can be seen from the difference in the lowest nuclear energy levels of the real and harmonic PES (black and green horizontal lines respectively). For comparison we show results for a dimer, where the number of PES are only nine and are well separated, for the interaction strengths $U=4$ (solid lines) and $U=1$ (dashed lines). The dissociation energy (the difference between minimum and asymptotic values of the lowest PES) is lower in the former case, which is also true for $L=5$. For the parameter regime we consider, the plasmon has minor impact on the PES structure, while crucially affecting the short-time dynamics. To gain further insight into the role played by the plasmon, we in the following section discuss the spectral and response functions.

\begin{figure}[t]
        \includegraphics[scale=0.55]{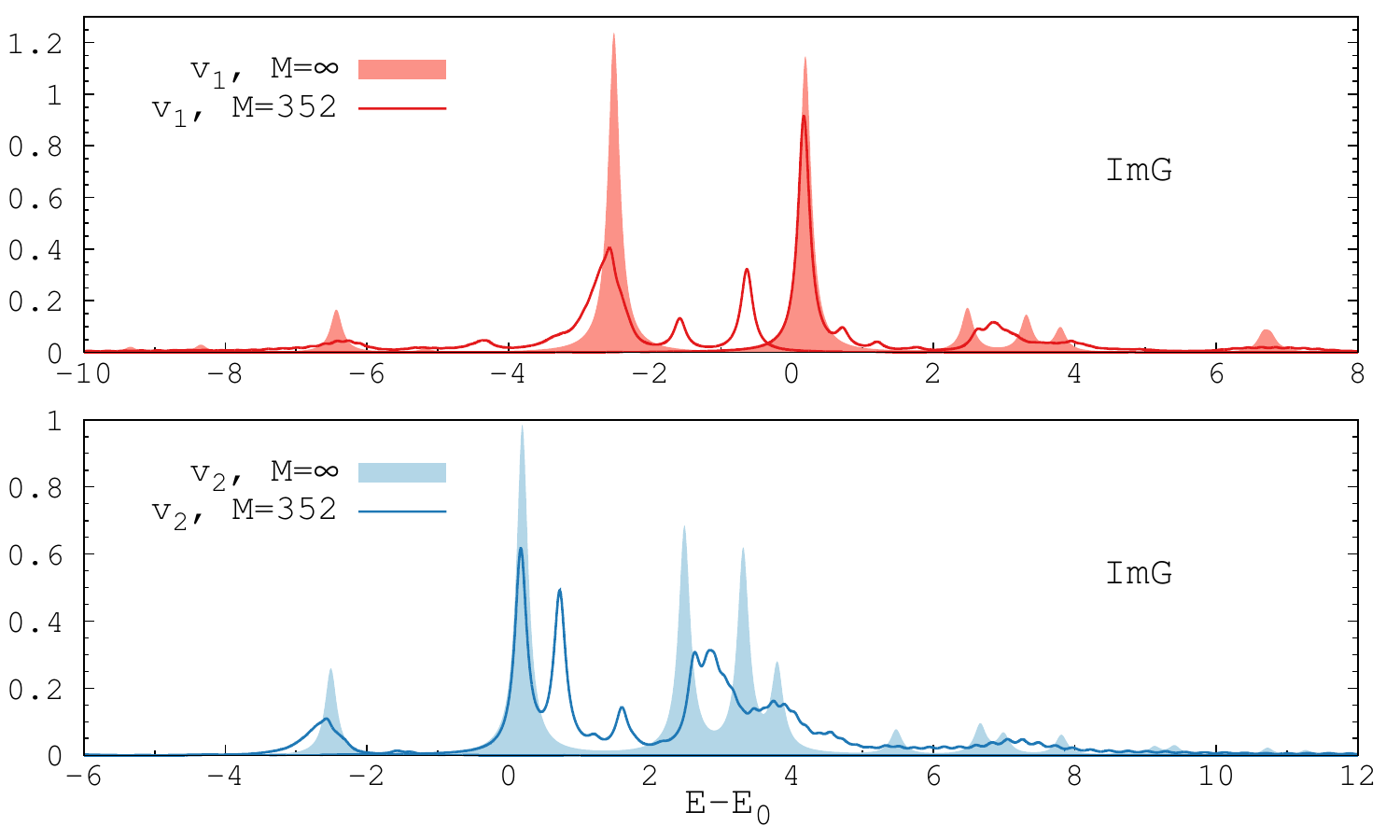}
\caption{The imaginary part of the local Green's function $G_{vv}$ is shown for the two valence levels $v_1$ (lower) and $v_2$ (upper). The shaded regions show the case of an infinitely heavy adsorbate and the lines correspond to a mass $M =352$, for a plasmon mode of frequency $\omega_p = 4$ and coupling $\gamma = 2$.}
\label{fig:langreth}
\end{figure}

\subsection{Spectral and response functions}
We here expand on the importance of plasmon effects in equilibrium, and to extract this information we look at the spectral functions $A(\omega)$ and $B(\omega)$ of the local one-particle Green function $G$ and density-density response function $\chi$ for the adsorbate levels. The zero-temperature Green function in the site basis is defined in equilibrium as
\begin{align}
G_{ij}(t) = \frac{1}{i}\langle \psi|T\{c_i(t)c^\dagger_j(0)\}|\psi\rangle
\end{align}
where $T$ is the time-ordering operator. Taking the Fourier transform of this expression we find the spectral function to be $A(\omega) = 2i\sign(\omega)\Im G(\omega)$. Similarly the density-density response function is defined by
\begin{align}
\chi_{ij}(t) = \frac{1}{i}\langle \psi|T\{\Delta \hat{n}_i(t)\Delta \hat{n}_j(0)\}|\psi\rangle,
\end{align}
where $\Delta \hat{n}_i = \hat{n}_i -n_i$ is the local density fluctuation operator, and from which the spectral function $B(\omega) = 2i\sign(\omega)\Im \chi(\omega)$ can once again be found by Fourier transform.

The electron-plasmon interaction term can be removed from the Hamiltonian via a Lang-Firsov transformation \cite{LangFirsov}, with the effect of renormalizing the values of the onsite energy and the electron-electron interaction. More specifically, we expect shifts of the sort $\epsilon_v \to \epsilon_v - \gamma^2/\omega_p$ and $U_{vv'} \to U_{vv'} - 2\gamma^2/\omega_p$. These features can both be observed in Fig.~\ref{fig:spectral} above, where the interaction induced gap in $\Im \chi$ is diminished, and the single-particle levels in $\Im G$ are shifted towards lower energies.

For $\gamma=2$, $\Im G$ shows several distinct peaked features that, loosely speaking, can be seen as emerging from broadened plasmon satellites \cite{Langreth70,Cini}, where the broadening is largely due to the mobility of the adsorbate and the inherent fluctuations. 
To support this argument, in Fig.~\ref{fig:langreth} we compare the density of states for an infinitely heavy adsorbate to a system with $M  = 352$. As soon as the mass becomes finite, i.e. the adsorbate mobility and the position fluctuations are increased, each peak is broadened and split into several smaller ones, and the weight of the distribution is shifted towards higher energies.

%
%%%%%%%%%%%%%%%%%%%%%%%%%%%%%%%%%%%%%%%%%%%%%%%%%%%
%%%%%%%%%%%%%%%%%%%%%%%%%%%%%%%%%%%%%%%%%%%%%%%%%%%
%%%%%%%%%%%%%%%%%%%%%%%%%%%%%%%%%%%%%%%%%%%%%%%%%%%
%%%%%%%%%%%%%%%%%%%%%%%%%%%%%%%%%%%%%%%%%%%%%%%%%%%
%%%%%%%%%%%%%%%%%%%%%%%%%%%%%%%%%%%%%%%%%%%%%%%%%%%
%%%%%%%%%%%%%%%%%%%%%%%%%%%%%%%%%%%%%%%%%%%%%%%%%%%
%%%%%%%%%%%%%%%%%%%%%%%%%%%%%%%%%%%%%%%%%%%%%%

\begin{figure}[t]
\includegraphics{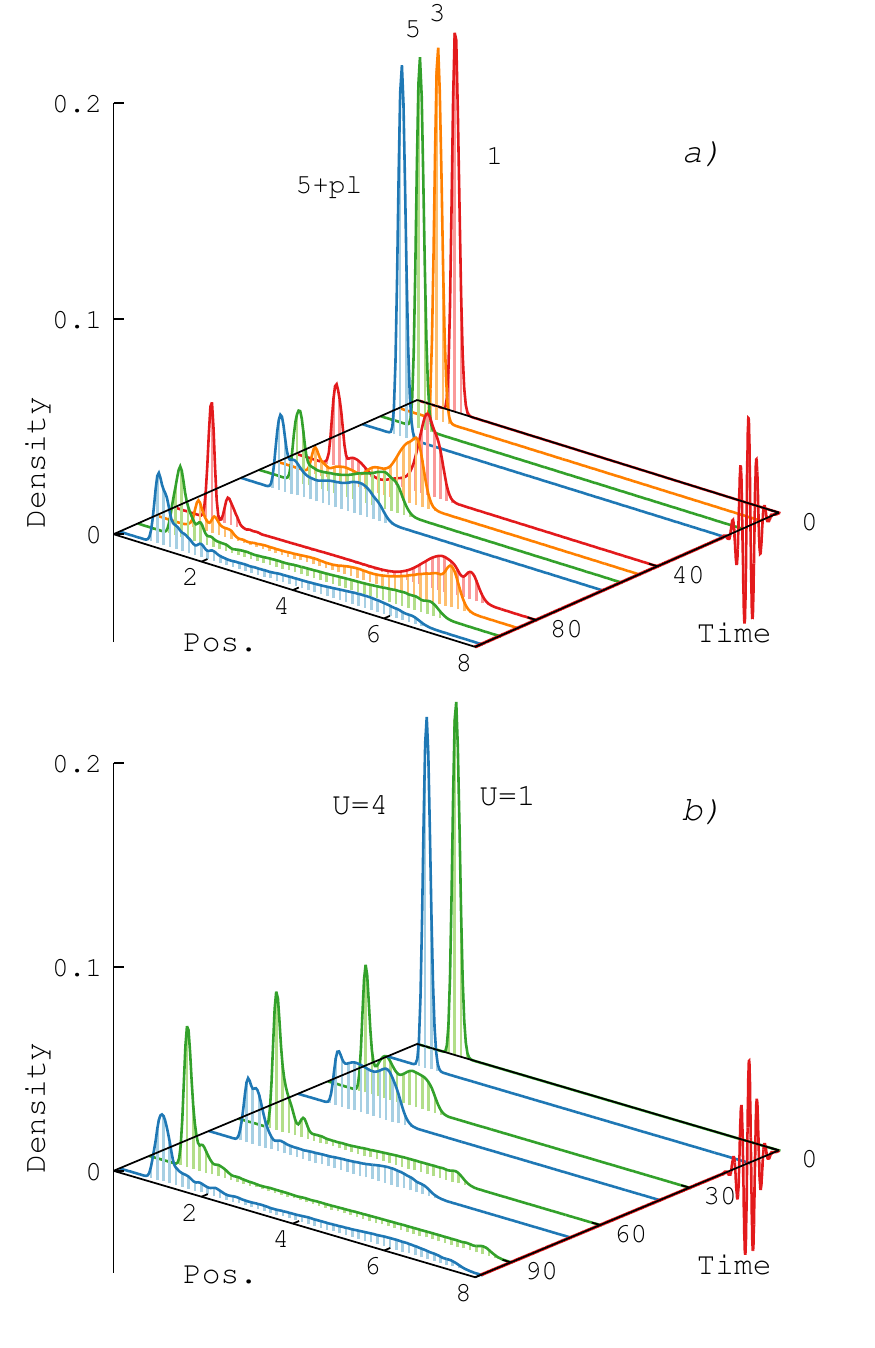}
\caption{Dependence of the adsorbate wavepacket dynamics on the length $L$ of the substrate and the interaction strength $U$. Panel $a$) shows snapshots of the wavepacket for $L=1$ (red) and $L=3$ (orange), as well as for $L=5$ with an electron-plasmon coupling $\gamma=0$ (green) and $\gamma = 1$ (blue).
In $b$) we show the evolution of the nuclear wavepacket for $U=1$ (green) and $U = 4$ (blue). In both panels $\Lambda_{vv'}(t)= Ae^{-(t- t_0)^2/\tau}\cos(\omega t)$, with $t_0 = 10$, $\tau = 13$ and $\omega = 6\pi/8$, of amplitude $A = 2$.}
\label{fig:sizeinteraction}
\end{figure}

\begin{figure}[t]
\includegraphics[scale=1]{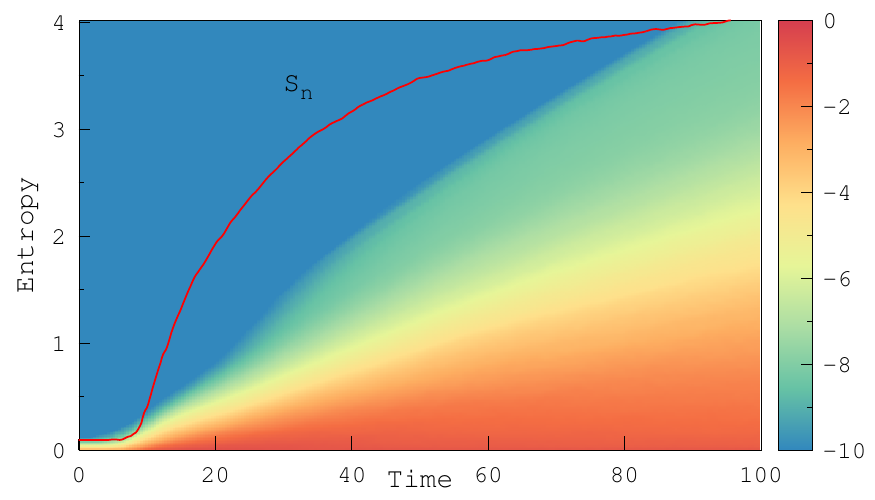}
\caption{Heat map of the occupation in the hundred first natural orbitals of the reduced nuclear density matrix, as a function of time. Superimposed is the entanglement entropy $S_n$ for the nuclear subsystem.}
\label{fig:orbitals}
\end{figure}

\section{Desorption dynamics}\label{sec:desorption}
Starting from $|g\rangle$, for $t>0$ the exact many-body wavefunction is
time evolved via the short iterated Lanczos algorithm \cite{ParkandLight}. 
To induce desorption dynamics we apply single- or double-pulse fields, and look at
the electron density $n_v(t)$ at the adsorbate level $v$, the mean internuclear position $x(t)$ and the nuclear probability distribution $P_t(x_k)$. The pulses we consider have a FWHM of 6fs (30fs), and a carrier wavelength of 800nm corresponding to a photon energy $\omega \simeq 1.5$ eV.

\subsection{Dependence on size and interaction strength}
To assess the role of the substrate on the adsorbate dynamics, in Fig.~\ref{fig:sizeinteraction}a
we show the adsorbate wavepacket $P_t(x_k)$ for substrates of length $L=1$, $3$ and $5$ and vanishing electron-plasmon coupling (red, orange and green curves). After the pulse has been applied,
$P_t(x_k)$ progressively spreads over larger internuclear distances. For $L=1$ the wavepacket is clearly split into two separate structures,
while for larger $L$ it is more uniformly distributed.
We define a desorption yield according to $Y(t) =1-\int_{0}^{x_0} P_t(x) dx$, where $x_0$ is chosen as the smallest value for which $Y(0) = 0$. At large times $Y= 0.65$, $0.85$ and $0.63$ for $L=1, 3$ and $5$ respectively, and a non-monotonic $Y$ appears to
be a rather general feature; we also observed it for much longer, non-interacting chains in the Ehrenfest approximation.
In the dipole approximation $Y$ appears to be only mildly sensitive to $L$. This is partly due to the system approaching the surface molecule limit ($\langle ge^{-\lambda (\hat{x}-1)}\rangle \simeq 1.8 > V$), and partly because the excitations induced by $\Lambda$ are within the adsorbate: we observe a stronger dependence of $Y$ on $L$ for fields coupling to the valence level density (not shown here).  This suggest desorption scenarios where non-local effects from the substrate play only a small role.
For $L=5$, we also include the local plasmon. The behaviour with and without plasmons (blue and green curves respectively) is about the same; however, with the plasmon, $Y = 0.59$ indicative of less desorption within the considered time-interval.

Electronic interactions are expected to have a non-negligible effect on desorption. To illustrate their role (Fig.~\ref{fig:sizeinteraction}b), we compare the distributions
$P_t(x_k)$ for $U=1$ and 4 that give the respective desorption yields $Y= 0.42$ and 0.59. For $U=1$ the adsorbate electronic density (not shown) 
fluctuates more in time, while for large interaction these oscillations are quenched. Thus the suppression of charge fluctuations to/from the substrate by electronic correlations appears to affect the probability of desorption.

\begin{figure}[thb]
        \includegraphics[scale=1]{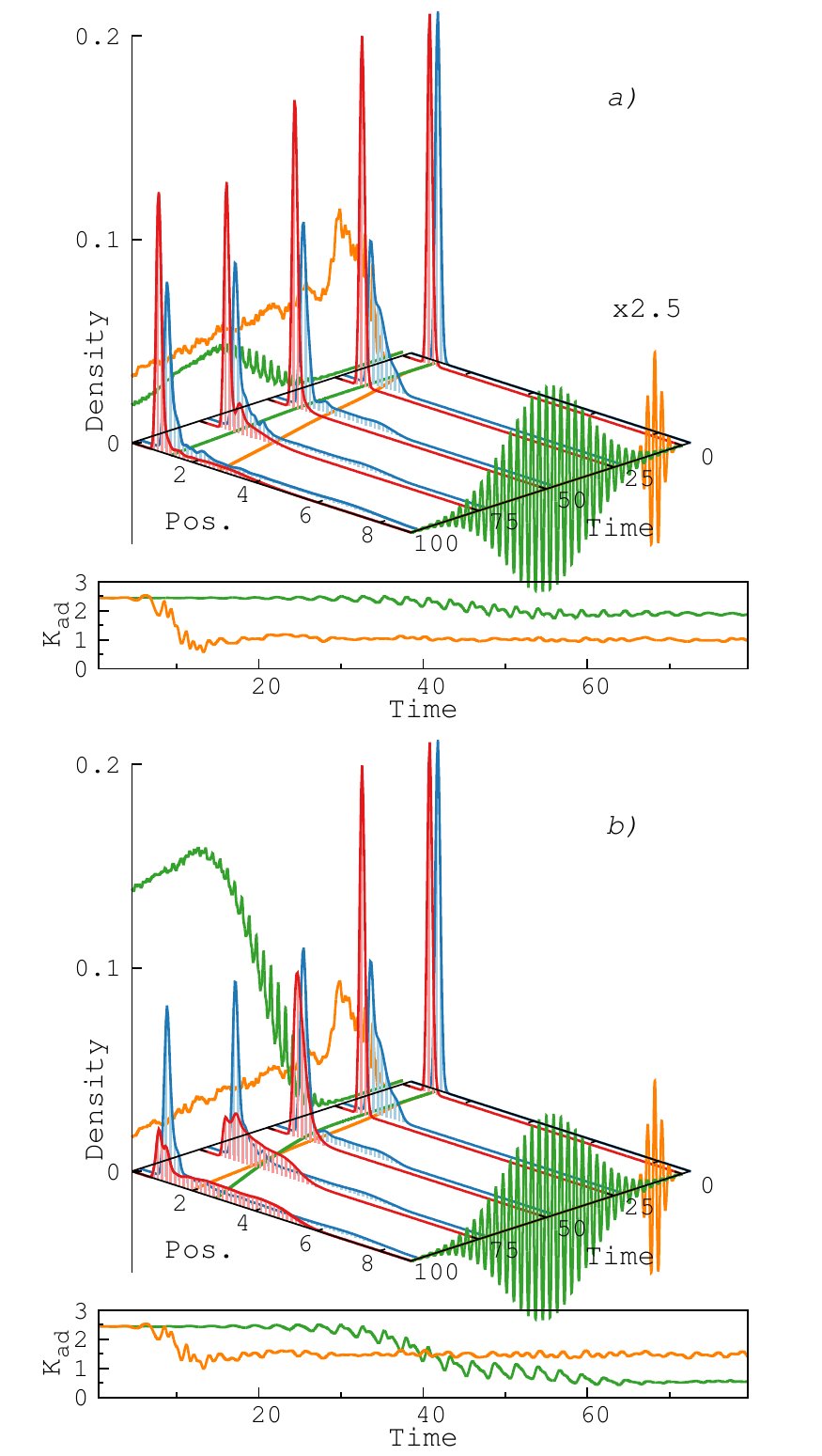}
\caption{Time evolution (in units of fs) of the nuclear density (red/blue), average position (orange/green) and plasmon density (back), for different pulse protocols (front). The parameters are as discussed in Section~\ref{sec:parameter} with $L=5$ and $U=4$. In $a$) two pulses of identical integrated intensity and duration $6$ fs (orange) and $35$ fs (green) are shown; in $b$) two pulses of identical amplitude $A = 2$ and duration $6$ fs (orange) and $35$ fs (green) are shown. The panels at the bottom show the respective bond kinetic energies $K_{ad}$.}
\label{fig:pulses}
\end{figure}

\subsection{Natural nuclear orbitals}
Using an adequate space grid can be a computational bottleneck for calculations addressing
desorption; furthermore, differently from on-resonance experiments, ultrashort pulses involve
a large spectrum of frequencies, and many PES are simultaneously involved. But how many is "many"?  
As an empirical answer, the heatmap in Fig.\ref{fig:orbitals} shows the occupation of the 
natural orbitals of the reduced nuclear density operator $\hat{\Gamma}(t) = \tr_{e,pl}\hat{\rho}(t)$, after the application of a 6fs pulse of amplitude $A = 3$.
The other system parameters are the same as for the blue curves in Fig. \ref{fig:sizeinteraction}b,
Here the trace is taken over the electronic and plasmonic degrees of freedom.
According to these results, a calculations would only require 
the first 50 or so natural orbitals to keep significant precision, reducing by a factor of 20 the currently used space grid basis,
but still maintaining full correlation between nuclear and electronic degrees of freedom. The entity of these correlations
can be easily realized when looking at the nuclear entanglement entropy $S_n=\tr \hat{\Gamma}\log \hat{\Gamma}$, which 
grows quickly already in the very early stages of desorption.

\begin{figure}[t]
\includegraphics[scale=1]{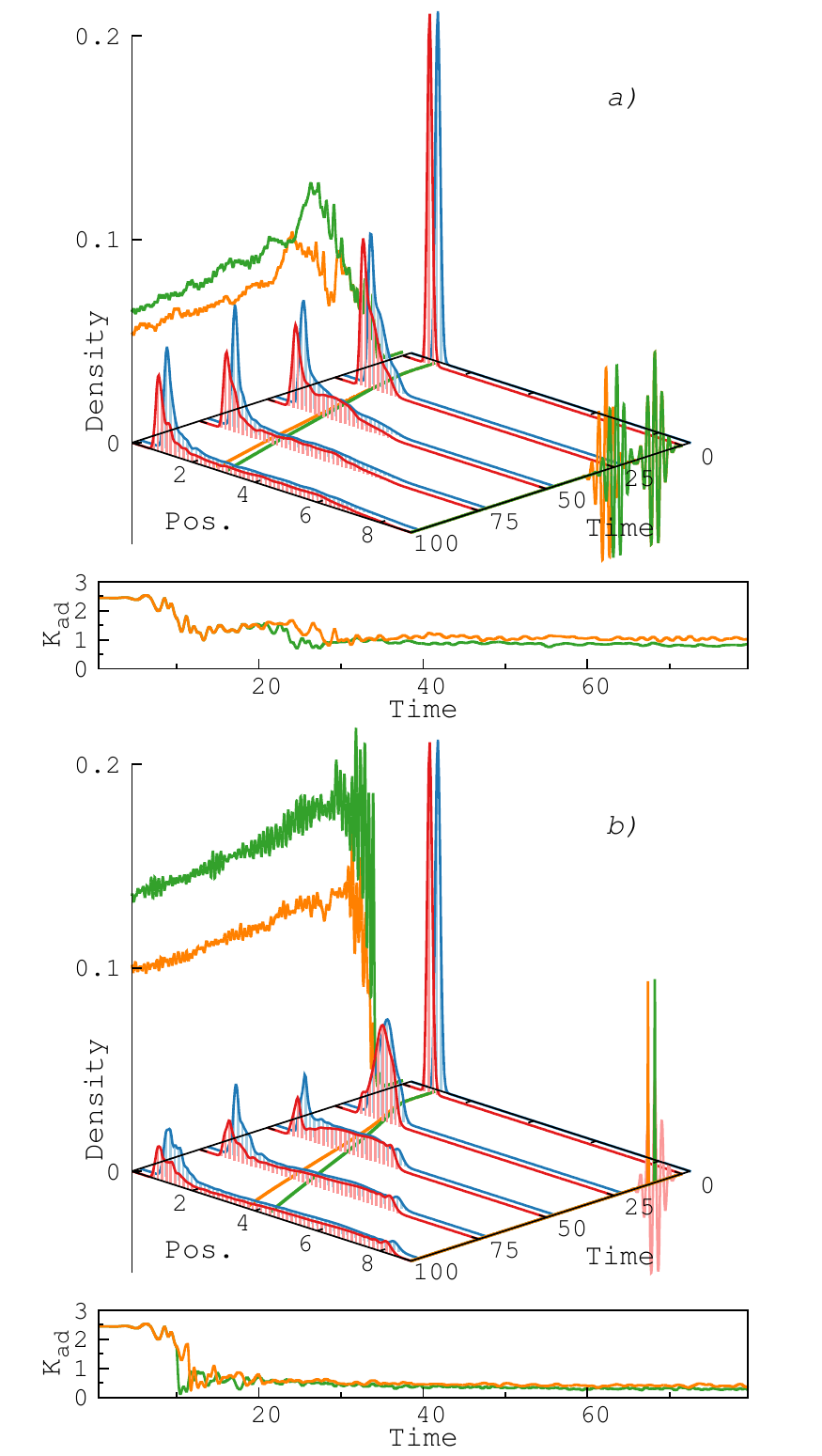}
\caption{Time evolution of the nuclear density (red/blue), average position (orange/green) and plasmon density (back), for two different pulse protocols (front). Parameters are as in Fig.~\ref{fig:pulses}. In $a$) two pulses are applied consecutively with delays of $14$fs or $18$fs, and $\Lambda_{vv'}(t)= Ae^{-(t- t_0)^2/\tau}\cos(\omega t) +Ae^{-(t- t_1)^2/\tau}\cos(\omega t+\varphi)$. Here $t_0 = 10$, $\tau = 13$, $\omega = 6\pi/8$ and $A=2$, with $t_1 = 24$ (red/green) or $t_1 = 28$ (blue/orange), and $\varphi$ chosen to give the second pulse the same envelope-carrier relation as the first. 
In $b$) a core electron is removed during the action of a 6fs pulse, at $\tau_c = 10$ (red/green) or $\tau_c = 11.3$ (blue/orange) with $w=6$.}
\label{fig:control}
\end{figure}

\subsection{Manipulating the system via pulse control}
We now go on to discuss the dynamics of a surface-adsorbate system induced by different pulse protocols, and to highlight the different time-scales at play we start by discussing a very simple
experiment where either the pulse duration or amplitude is varied. For this purpose we find it convenient to analyze the behavior of the the bond kinetic energy $K_{as} = \sum_v \left(c_v^\dagger c_s + c_s^\dagger c_v\right)$, which we take as our measure of the adsorbate-surface bond strength.

In Fig.~\ref{fig:pulses}a we compare two 800nm (IR) pulses with FWHMs of 6fs and 35fs, and of equal integrated intensity. In the first case we see a significant amount of plasmon excitation and reduction of the bond kinetic energy between the surface and the adsorbate. In the second case there is much less change in both plasmon density and kinetic energy, leading also to much less desorption. Since the only difference is the duration of the pulse the different outcomes are most likely due to the fact that, for the shorter pulse, non-adiabatic effects in the response 
play a greater role.

In Fig.~\ref{fig:pulses}b we instead compare two pulses of equal amplitude $A=2$, but with with FWHMs of 6fs and 35fs. The latter leads to a much larger displacement of the adsorbate within the same time-frame, as expected from its greater (by a factor 2.5) deposited energy. For the 35fs pulse the evolution of the system is dominated by the shape of the field, while for the 6fs pulse its behaviour is to a larger extent determined by the internal dynamics. While this is a very reasonable physical result, it also points to the observation that ultrafast dynamics within the first 50 fs occurs both for the electron and nuclear dynamics in a correlated fashion, and thus separating these two time scales might not be appropriate. This kind of correlation is experimentally observed for molecular systems \cite{Zhou}.

Finally, we use the model to gain insight into two prototypical experiments, based on pulse durations and wavelengths realizable with existing state-of-the-art lasers \cite{laser,Huillier,Marsell2}. In Fig.~\ref{fig:control}a we apply in a pump-probe manner \cite{Hornett} two 6fs pulses with delays differing by $1.5$ IR cycles, to explore the possibility of coherent control of the adsorbate motion. The effect of changing the delay is clear: while both cases give an increase in the average adsorbate position, the shorter delay leads to a $15$ percent larger stretching (after $100$fs). 
This is due to the plasmon: in contrast to the equilibrium case, it acts as a strong harmonic perturbation which,
during its cycle, can be reinforced by applying a second pulse at the right point \cite{Linic,Gadzuk}. 

As a second example, in Fig.~\ref{fig:control}b an instantaneous core-level photoemission (PE) is combined with an IR pulse, as conducted in IR+XUV experiments using high harmonic generation technology \cite{Ivanov}. Ejecting an electron has a huge impact, and the average position of the adsorbate in time is almost four times larger compared to only an IR pulse (see Fig.~\ref{fig:pulses}b). Interestingly, a change of the PE time with less than an IR cycle will significantly influence both the adsorbate dynamics and the plasmon behaviour.
As an overall, final observation about Fig.~\ref{fig:control}, it can also be seen that while the moderate field strengths used here do not lead to desorption within the first tenths of fs (as found in some molecular systems \cite{Zhou,Neutze}), significant bond-stretching occurs so rapidly that separating nuclear and electronic timescales may not be justified for these experiments.

\begin{figure}
\includegraphics{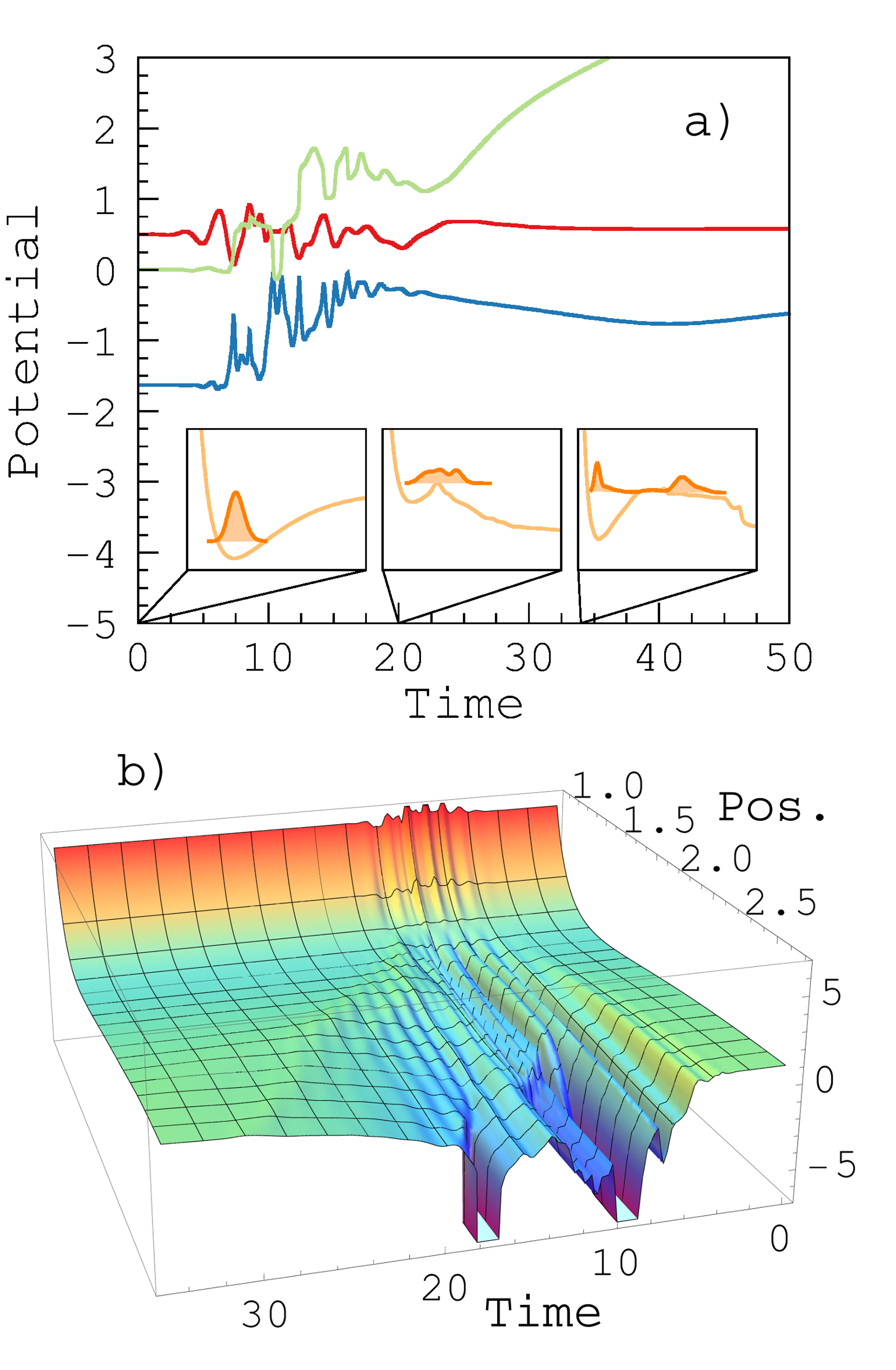}
\caption{Panel $a$) gives the exact nuclear potential $\epsilon_{KS}$ for a dimer with one valence level at $\epsilon_1 = -U/2$, while
$b$) shows the electron density at the adsorbate (red) and the exact electronic KS potential $|T_{KS}|$ (blue) and $\arg (T_{KS})$ (green) for the same system. The insets are snapshots at $t=0$ $t=17$ and $t=34)$ of the nuclear wavepacket and corresponding $\epsilon_{KS}$. In all cases $\Lambda_{vv'}(t)= Ae^{-(t- t_0)^2/\tau}\cos(\omega t)$, with $t_0 = 10$, $\tau = 13$ and $\omega = 6\pi/8$, of amplitude $A = 10$.}
\label{fig:TDDFT}
\end{figure}

\section{A TDDFT perspective}\label{sec:tddft}
As a way to obtain insight into the desorption dynamics we consider a multi-component TDDFT approach \cite{LiTong86,Butriy},
specialised to electrons on a lattice. A system with electrons on a set of spin-orbitals 
$\{i,\sigma_i\}$ in a lattice, and
nuclei at $\{R_m\} \equiv \{{\bf R}_m,\zeta_m\}$ (where ${\bf R}_m$ and $\zeta_m$ denote space and spin variables respectively), can be described by a Hamiltonian with external potentials $T_{ij}^{ext}(t)$ and $\epsilon^{ext}({\bf R},t)$. On the nuclear side, we choose as fundamental variable the diagonal $\Gamma({\bf R},t)$ of the one-particle density matrix \cite{Butriy}. For the electrons we observe that increasing the internuclear distance should result in a reduced hopping probability, so the KS Hamiltonian $H_{KS}$ should in the lattice basis have (in general complex) matrix elements $T^{KS}_{ij}(t)$ where both modulus and phase can vary. This is taken into account by a generalization of a lattice time-dependent current DFT that uses
the complex bond current as basic variable \cite{Tokatly11}, defined as $Q_{ij}^\sigma(t) = T_{ij}^{ext}(t)\rho_{ij}^\sigma(t) + \tilde\rho_{ij}^\sigma(t)$. For a purely electronic system only $\rho_{ij}^\sigma(t)=\langle \psi(t)| a^\dagger_{i\sigma}a_{j\sigma}|\psi(t)\rangle$ would enter, and the explicit electron-nuclear coupling $V_{ij}(\{\hat{\bf R}_m\})$ is contained in the second term $\tilde{\rho}_{ij}^\sigma(t) =\langle \psi(t)| V_{ij}(\{\hat{\bf R}_m\})a^\dagger_{i\sigma}a_{j\sigma}|\psi(t)\rangle$ (there is of course an implicit dependence through the state vector).
Proceeding as in \cite{Tokatly11,Farzhanepour14} a bijective mapping between the fundamental variables ($Q_{ij},\Gamma$) and the potentials ($T_{ij}^{ext},\epsilon^{ext}$)
can be established \cite{Bostrompreparation}, and as discussed in \cite{Tokatly11} $T$-representability (and 
non-interacting $T$-representability) by $T^{KS}_{ij}$ is ensured when $|\rho_{ij}(t)| > 0$. 

\begin{figure*}
\hspace{-1cm}
    \begin{minipage}{.4\textwidth}
        \centering
        \includegraphics[scale=0.95]{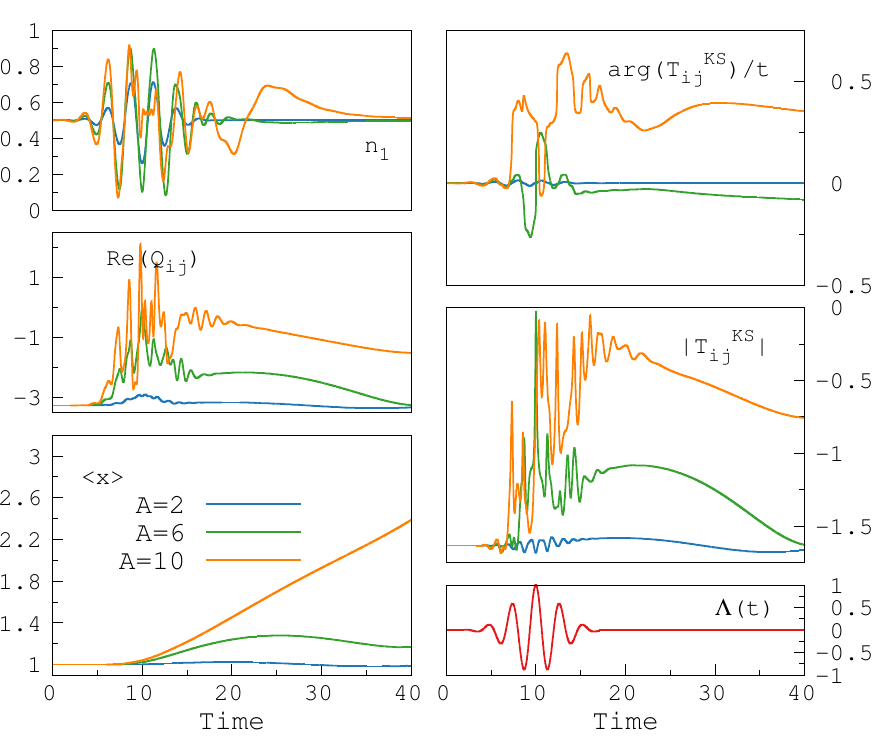}
    \end{minipage}%
\hspace{2cm}
    \begin{minipage}{.4\textwidth}
        \centering
    \includegraphics[scale=0.16]{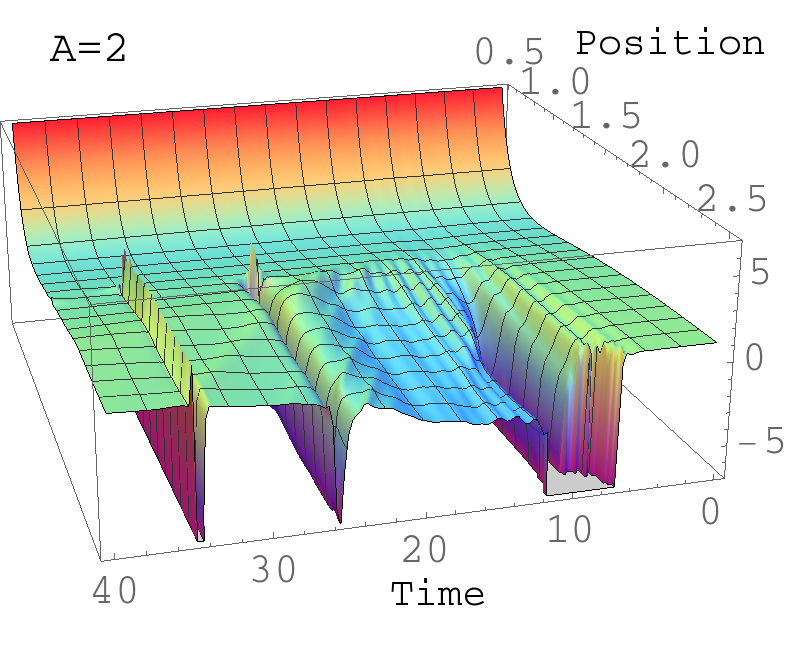}

    \includegraphics[scale=0.16]{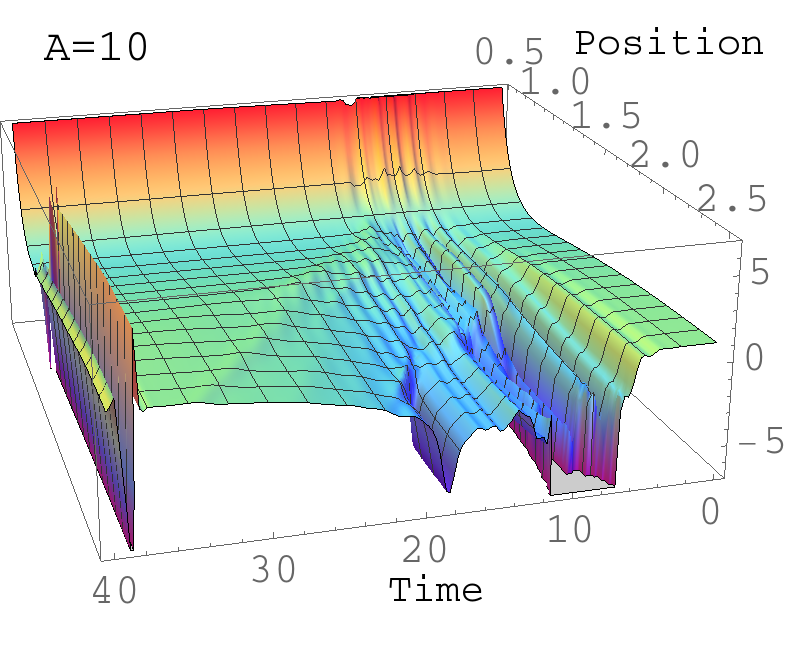}
    \end{minipage}
\caption{In the three leftmost panels we show the electron density, bond kinetic energy and mean internuclear distance, after the application of a square potential with amplitude $A = 2$ (blue), $A = 6$ (green) or $A = 10$ (orange). The three central panels show the argument and modulus of the exact electronic Kohn-Sham potential $T^{KS}(t)$, as well as the time structure of the external laser field. In the panels furthest to the right we show 3D plots of the exact nuclear potential for pulses of amplitude $A = 2$ (top) and $A = 10$ (bottom).}
\label{fig:fast}
\end{figure*}

We show in Fig.~\ref{fig:TDDFT} results for a dimer ($L=1$) without plasmons ($\gamma=0$), with one valence level on the adsorbate, and write ${\bf R} \to x$ for the single, 1D nuclear coordinate of our model. For the system considered the Kohn-Sham Hamiltonians for the electronic and nuclear system are respectively
\begin{align}\label{eq:H_e}
\hat{H}_e^{KS} =& \sum_{ij,\sigma}\left(T_{ij}^{KS}[Q_{ij},\Gamma](t)c_{i,\sigma}^\dagger c_{j,\sigma} + h.c.\right)
\end{align}
and
\begin{align}\label{eq:H_n}
\hat{H}_n^{KS} =& \sum_{k}\frac{p^2}{2m} + \epsilon_{KS}[Q_{ij},\Gamma](x,t).
\end{align}
and this simplified case (a purely electronic Hubbard dimer is non-interacting $v$-representable \cite{Verdozzi2008}) already shows essential features of the KS potentials $T^{KS}$ and $\epsilon_{KS}$, which in fact can be constructed exactly.

For the electrons, the complex $T^{KS}_{ij}(t)$ are determined via numerical reverse engineering. 
To determine $\epsilon_{KS}(x,t)$ we perform an exact factorisation of the wavefunction of the interacting electron-nuclear system \cite{Gross}, by defining a nuclear wavefunction $\chi(x,t) = e^{-iS(x,t)}\xi(x,t)$. Choosing the gauge where the vector potential is zero \cite{Gross}, the exact nuclear potential is $\epsilon =(2M)^{-1}[ \left(\partial_x \ln \xi \right)^2 + \partial_{xx}\ln \xi - \left(\partial_x S\right)^2] - \partial_t S$. This is also the exact KS nuclear potential ($\epsilon_{KS}(x_k,t)$, after introducing a discrete mesh $x_k$ for the nuclear coordinate).

In Fig.~\ref{fig:TDDFT}a the splitting of the nuclear wavepacket during desorption is seen to reflect in the behaviour of $\epsilon_{KS}(x_k,t)$: for $t\simeq 10$ and 17, $\epsilon_{KS}$ develops dips to dynamically push outwards part of the wavepacket. Between these times it undergoes several dynamical corrugations, related to the electronic oscillations in turn induced by the external pulse. This is seen also in the snapshots at the bottom of Fig.~\ref{fig:TDDFT}b, showing the nuclear wavepacket and the corresponding $\epsilon_{KS}$.
The modulus $T^{KS}$ of the electronic KS potential decreases for large adsorbate-substrate distances
(as expected on physical grounds), while the phase appears to grow in a steady fashion; as a reference we show the adsorbate electronic density (red curve). During the central phase of the desorption process we see rapid oscillations in $T^{KS}$: on speculative grounds, such a non-trivial temporal pattern
suggests that, in general, desorption and charge transfer at surfaces can be quite challenging to a TDDFT description \cite{Fuks}, e.g. when adiabatic TDDFT treatments are considered.

\begin{figure*}
\hspace{-1cm}
    \begin{minipage}{.4\textwidth}
        \centering
        \includegraphics[scale=0.95]{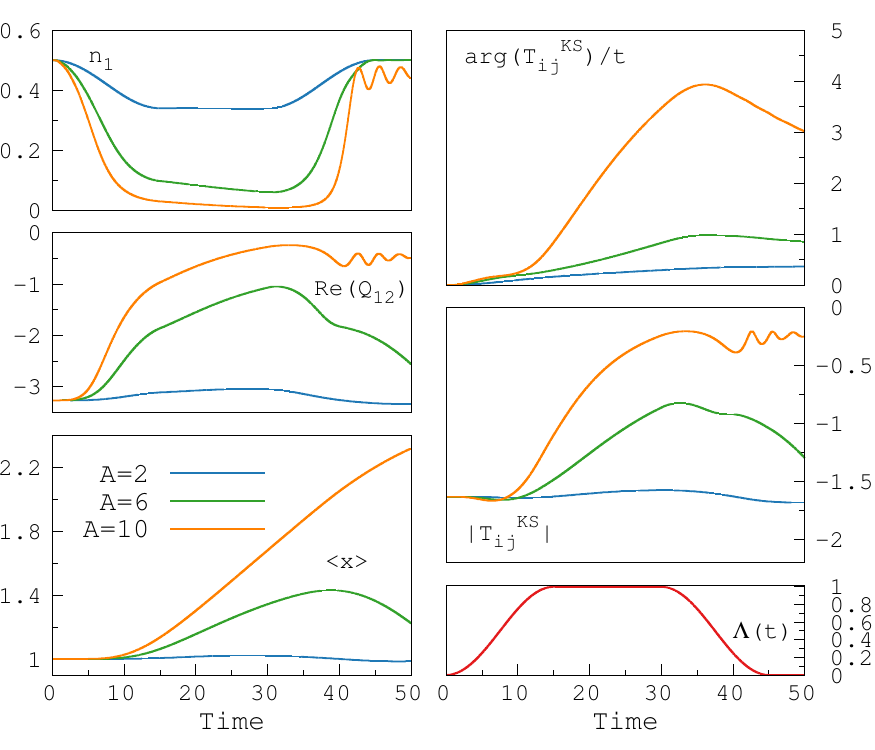}
    \end{minipage}%
\hspace{2cm}
    \begin{minipage}{.4\textwidth}
        \centering
    \includegraphics[scale=0.16]{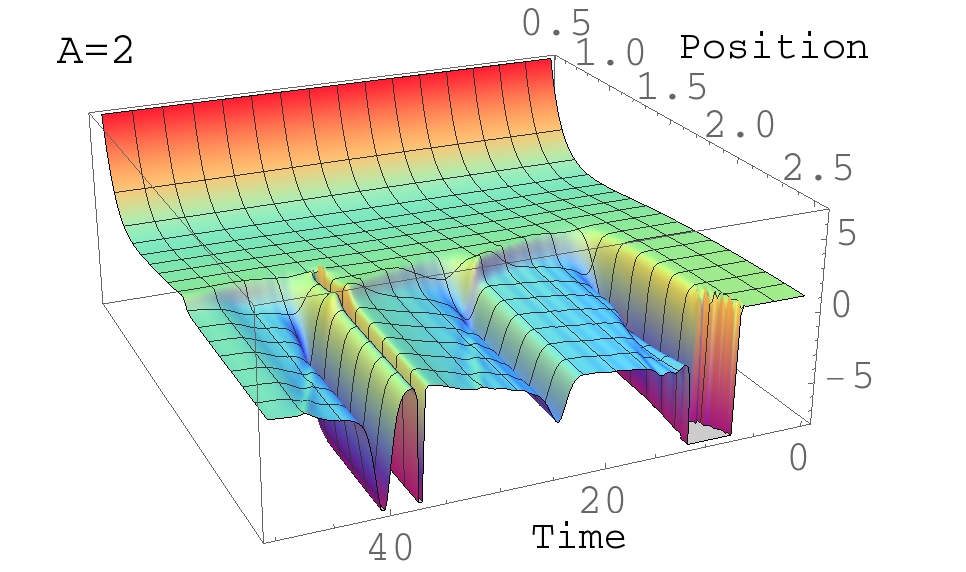}

    \includegraphics[scale=0.16]{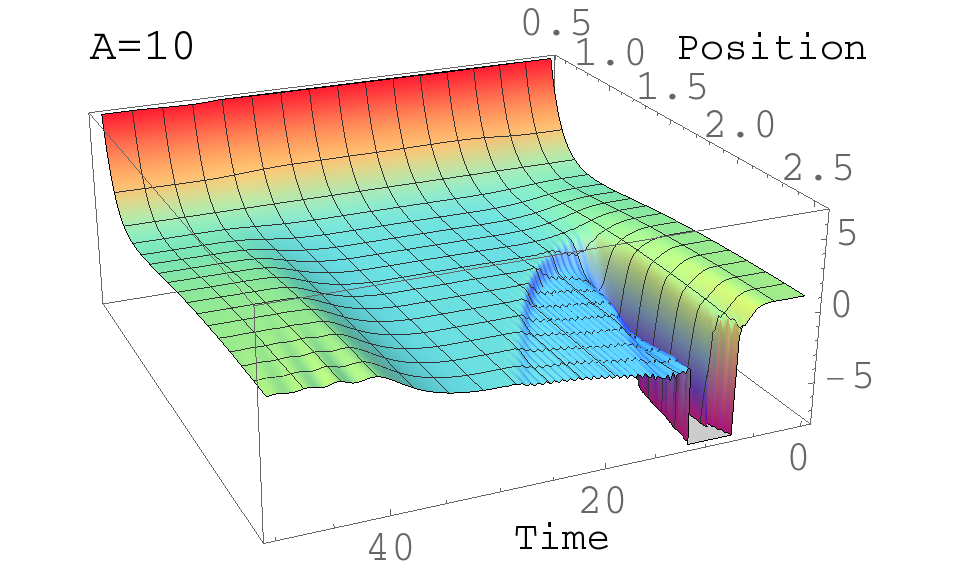}
    \end{minipage}
\caption{In the three leftmost panels we show the electron density, bond kinetic energy and mean internuclear distance, after the application of a square potential with amplitude $A = 2$ (blue), $A = 6$ (green) or $A = 10$ (orange). The three central panels show the argument and modulus of the exact electronic Kohn-Sham potential $T^{KS}(t)$, as well as the time structure of the external laser field. In the panels furthest to the right we show 3D plots of the exact nuclear potential for pulses of amplitude $A = 2$ (top) and $A = 10$ (bottom).}
\label{fig:intermediate}
\end{figure*}

\subsection{Adiabatic approximation}
The potentials $T_{ij}^{KS}$ and $\epsilon_{KS}$ can in general be very complicated, but there exist cases where an adiabatic approximation could be expected to perform rather well, even in the limit of significant desorption. In Fig.~\ref{fig:fast} we compare the expectation values of the electronic density, bond kinetic energy and mean internuclear position, with the external field given by a Gaussian pulse with FWHM $6$fs and carrier wavelength of $800$nm, and an amplitude of $2$ (blue), $6$ (green) and $10$ (orange) respectively. In all cases there are rapid oscillations in the electronic quantities, which are reflected in the KS potentials. Even in the case of weak perturbation where the adsorbate stays bound and performs oscillations (blue), we see features in the potentials that could be hard to reproduce in the adiabatic approximation.

\begin{figure*}
\hspace{-1cm}
    \begin{minipage}{.4\textwidth}
        \centering
        \includegraphics[scale=0.95]{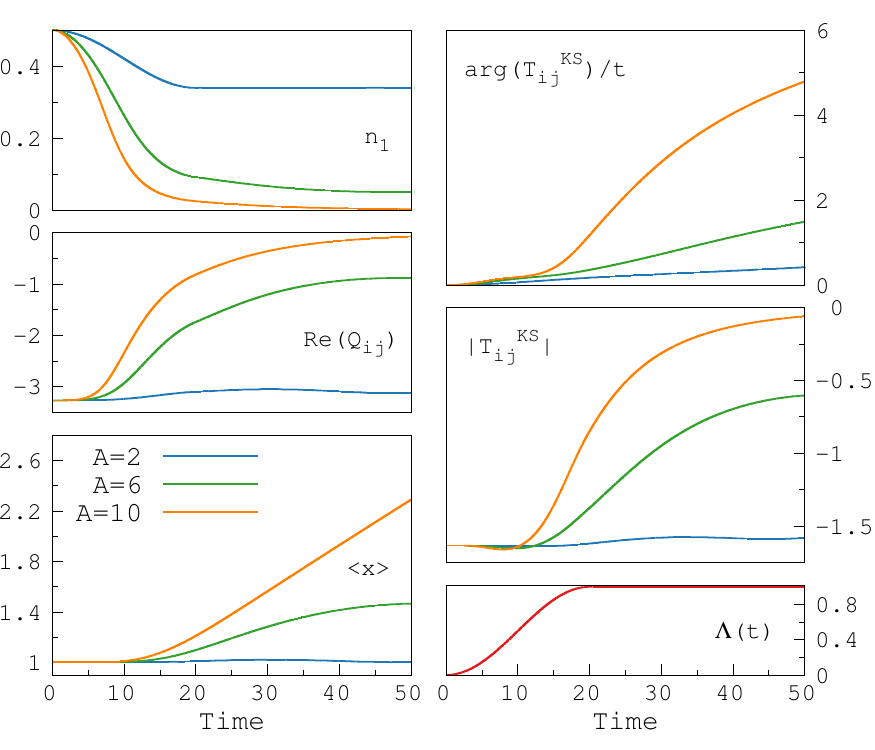}
    \end{minipage}%
\hspace{2cm}
    \begin{minipage}{.4\textwidth}
        \centering
    \includegraphics[scale=0.16]{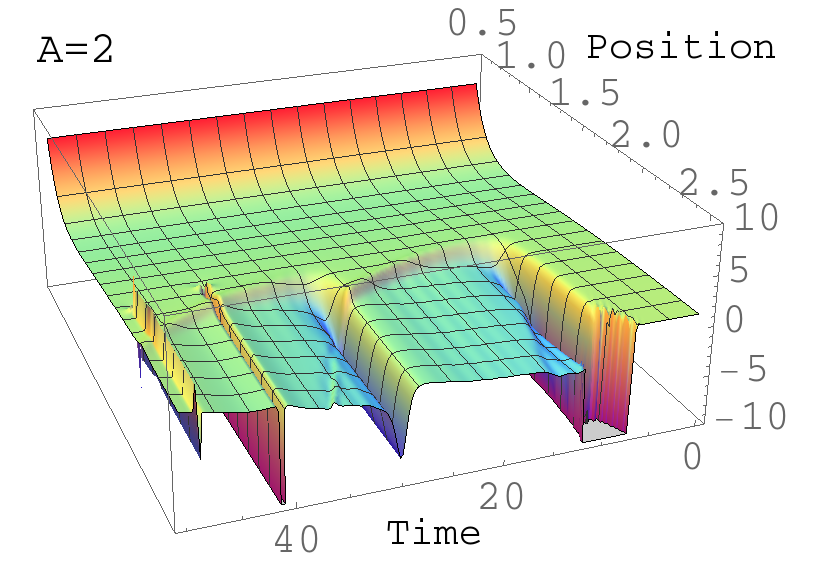}

    \includegraphics[scale=0.16]{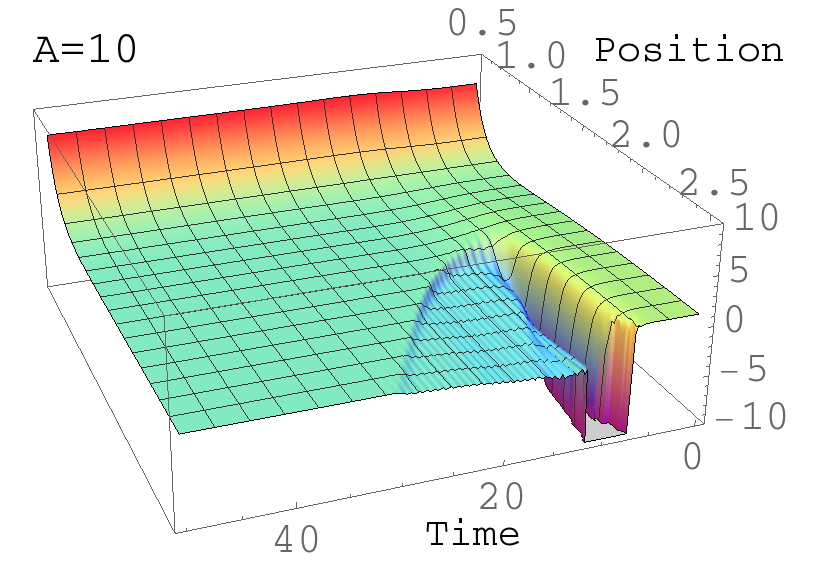}
    \end{minipage}
\caption{In the three leftmost panels we show the electron density, bond kinetic energy and mean internuclear distance, after the application of a square potential with amplitude $A = 2$ (blue), $A = 6$ (green) or $A = 10$ (orange). The three central panels show the argument and modulus of the exact electronic Kohn-Sham potential $T^{KS}(t)$, as well as the time structure of the external laser field. In the panels furthest to the right we show 3D plots of the exact nuclear potential for pulses of amplitude $A = 2$ (top) and $A = 10$ (bottom).}
\label{fig:slow}
\end{figure*}

In Fig.~\ref{fig:intermediate} we show the same quantities for a square-like external field, given by $\Lambda(t) = \chi_{[0,\tau]}(t)\sin^2\left(\pi t/2\tau\right) + \chi_{[\tau,2\tau]}(t) + \chi_{[2\tau,3\tau]}(t)\left[1-\sin^2\left(\pi t/2\tau-\pi\right)\right]$, where $\chi_I(t)$ is the characteristic function of the interval $I$ ($\chi_I(t) = 1$ for $t\in I$ and $0$ otherwise) and $\tau = 15$. In this case the behavior of the KS potentials is much smoother, especially during the action of the pulse. For the strongest perturbation, leading to a significant amount of desorption, the amplitude of the electronic potential shows rapid oscillations at the end of the pulse, in contrast to the phase which is well behaved for all times.

In Fig.~\ref{fig:slow} we show the case of a smeared step potential, given explicitly by $\Lambda(t) = \chi_{[0,\tau]}(t)\sin^2\left(\pi t/2\tau\right) + \theta(t-\tau)$ with $\tau = 20$. We see that all quantities change smoothly during the application of the pulse, independently of the strength of the perturbation. For the strongest external field we see clear indications of desorption for large times, and as expected the amplitude of the hopping parameter approaches zero in this limit. It is thus likely that an adiabatic approximation within TDDFT would perform well under these circumstances.

As a final comment we note that the nuclear KS potential is typically smoother for stronger perturbations, which can be understood from the evolution of the nuclear density distribution (see Figs.~\ref{fig:sizeinteraction}-\ref{fig:control}). For weak perturbations part of the density is ejected around the time that the pulse acts, while another part stays bound. The part that remains in the potential well performs oscillations, and each time it reaches the turning point of the potential part of the density is emitted. These successive emissions are reflected in the successive dips in the nuclear KS potential, that can be see for all three perturbations. For a strong perturbation there is only one emission, where the whole wavepacket is released, and after this the potential behaves quite smoothly.

%

%%%%%%%%%%%%%%%%%%%%%%%%%%%%%%%%%%%%%%%%%%%%%%%%%%

\section{Conclusions}\label{sec:conclusion}
In this work, we have introduced an exactly solvable model for electron-nuclear dynamics of adsorbates induced by ultrashort laser pulses. Though finite in size, the systems we can treat still contain the rich behaviour expected for an adsorbate-surface system. To illustrate the broad scope of the model, we briefly touched upon several issues, e.g. adsorbate dynamics in the surface molecule limit, electronic correlations and manipulation of plasmon dynamics. We also showed that the model can be a valuable aid in devising laser schemes to control the outcome of surface studies of light-matter interaction. Further, it can be used to benchmark more realistic theoretical approaches where approximations are necessarily introduced
(it was e.g. used here to gain insight into general features of the KS potentials of TDDFT during desorption).
As adsorbate systems are a key paradigm to explore light-matter interactions, and new laser sources will increasingly be applied to surfaces and thin films, work to extend and apply our model in several directions is under way.

\begin{acknowledgments}
We thank Carl-Olof Almbladh for fruitful discussions. This work was supported by Swedish (VR)  and European (ERC) Research Councils.
\end{acknowledgments}

%%%%%%%%%%%%%%%%%%%%%%%%%%%%%%%%%%%%%%%%%%%%%%%%%%%%%%%%%%%%
%%%%%%%%%%%%%%%%%%%%%%%%%%%%%%%%%%%%%%%%%%%%%%%%%%%%%%%%%%%%


\begin{thebibliography}{999}
%
\bibitem{Zewail} A. H. Zewail, J.\ Phys.\ Chem.\ A {\bf 104}, 5660 (2000).

\bibitem{FarisG} C. Miron {\it et al.}, Nature Physics {\bf 8} 135 (2012).

\bibitem{Henriksen} A. Garc\'{i}a-Vela and N. E. Henriksen, J. Phys. Chem. Lett. {\bf 6}, 824 (2015).

\bibitem{Gisselbrecht} E. P. M\aa nsson {\it et al.}, Nature Physics {\bf 10}, 207 (2014).

\bibitem{Remacle} F. Remacle {\it et al.}, J. Phys. Chem. A {\bf 103}, 10149 (1999).

\bibitem{PerStef} E. Perfetto {\it et al.}, Physical Review A {\bf 92}, 033419 (2015).

\bibitem{Frischkorn} C. Frischkorn and M. Wolf, Chem.\ Rev.\ {\bf 106}, 4207 (2006).

\bibitem{Petek1} H. Petek and S. Ogawa, Annu. Rev. Phys. Chem. {\bf 53}, 507 (2002).

\bibitem{Gabor} G. A. Somorjai and Y. Li, {\it Introduction to Surface Chemistry and Catalysis} (Wiley, 2010).

\bibitem{Hornett} S. M. Hornett {\it et al.}, Phys. Rev. B {\bf 90}, 081401(R) (2014).

\bibitem{Linic} S. Linic {\it et al.}, Nature Materials {\bf 14}, 567 (2015).

\bibitem{SurfScience} M. A. Henderson, Surf. Sci. Rep. {\bf 66}, 185 (2011).

\bibitem{Keller} U. Keller, Nature {\bf 424}, 831 (2003). 

\bibitem{Sutter} D. H. Sutter {\it et al.}, Opt.\ Lett.\ {\bf 24}, 631 (1999). 

\bibitem{Aeschlimann} M. Aeschlimann {\it et al.}, Science {\bf 333}, 1723 (2011).

\bibitem{Marsell} E. M\aa rsell {\it et al.}, Annalen der Physik {\bf 525}, 162 (2013). 

\bibitem{Bartels} L. Bartels {\it et al.}, Science {\bf 305}, 648 (2004).    

\bibitem{Backus} H. G. Backus {\it et al.}, Science {\bf 310}, 1790 (2005).    

\bibitem{KiHyun} Ki Hyun Kim {\it et al.}, Phys.\ Rev.\ Lett.\ {\bf 107}, 047401 (2011).   

\bibitem{Muino} R. D\'{i}ez Mui\~no {\it et al.}, Proc. Nat. Acad. Sci. U.S.A. {\bf 108}, 971 (2011).   

\bibitem{Fuchsel} G. Fuchsel {\it et al.}, Phys.\ Rev.\ Lett.\ {\bf 109}, 098303 (2012). 

\bibitem{Petek2} H. Petek, J.\ Chem.\ Phys.\ {\bf 137}, 091704 (2012).

\bibitem{Nuernberger} P. Nuernberger {\it et al.}, PCCP {\bf 14}, 1185 (2012).

\bibitem{Tremblay} G. Fuchsel {\it et al.}, Chem. Phys. Phys. Chem. {\bf 14}, 1471 (2013).

\bibitem{DellAngela} M. Dell'Angela {\it et al.}, Science {\bf 339}, 1302 (2013).      

\bibitem{BalzerBonitzbook} K. Balzer, and M. Bonitz, {\it Lecture notes in Physics}, {\bf 867}, Springer (2013).
 
\bibitem{bookNEG} G. Stefanucci and R. van Leeuwen, {\it Nonequilibrium Many-Body Theory of Quantum Systems} (Cambridge University Press, 2013).

\bibitem{RG84} E. Runge and E. K. U. Gross, Phys. Rev. Lett. {\bf 52}, 997 (1984).

\bibitem{EsaRas} A. Crawford-Uranga {\it et al.}, Phys. Rev. A. {\bf 90}, 033412 (2014).

\bibitem{Hellgren} M. Hellgren, E. R\"as\"anen, and E. K. U. Gross, Phys. Rev. A {\bf 88}, 013414 (2013).

\bibitem{Gross} A. Abedi, N. T. Maitra, and E. K. U. Gross, Phys.\ Rev.\ Lett.\ {\bf 105}, 123002 (2010).

\bibitem{RvL2}N. S\"akkinen {\it et al.}, arXiv:1403.2968.

\bibitem{Butriy} O. Butriy {\it et al.},  Phys. Rev. A {\bf 76}, 052514 (2007).

\bibitem{Tavernelli} E. Tapavicza, I. Tavernelli, and U. Rothlisberger,  Phys. Rev. Lett. {\bf 98}, 023001 (2007).

\bibitem{Prezhdo} C. F. Craig, W. R. Duncan, and O. V. Prezhdo, Phys. Rev. Lett. {\bf 95}, 163001 (2005).

\bibitem{dipole} N. E. Dahlen and R. van Leeuwen, Phys. Rev. Lett. {\bf 98}, 153004 (2007).

\bibitem{Stefanucci} E. Perfetto and G. Stefanucci, Phys. Rev. A {\bf 91}, 033416 (2015).

\bibitem{Perfetto1} E. Perfetto, D. Sangalli, A. Marini, and G. Stefanucci, arXiv:1507.01786v1.
  
\bibitem{Anderson} P. W. Anderson, Phys. Rev. {\bf 124}, 41 (1961).

\bibitem{Newns} D. M. Newns, Phys. Rev. {\bf 178}, 1123 (1969).

\bibitem{Grimley} T. B. Grimley, Proc. Phys. Soc.  {\bf 90}, 751 (1967).

\bibitem{SchonGun} K. Sch\"onhammer and O. Gunnarsson, Solid State Commun. {\bf 23}, 691 (1977); {\bf 26}, 147 (1978).
  
\bibitem{Langreth70} D. Langreth, Phys. Rev. B {\bf 1}, 471 (1970).

\bibitem{KotaToyo} A. Kotani and Y. Toyozawa, J. Phys. Soc. Jpn. {\bf 35}, 1073 (1973).

\bibitem{Gumhalter} B. Gumhalter, J. Phys. France {\bf 38}, 1117 (1977). 

\bibitem{CiniSS79}  M. Cini, Surf. Sci. {\bf 79}, 589 (1979).
  
\bibitem{Sugano} K. Shinjo and S. Sugano, Phys. Rev. B {\bf 30}, 604 (1984).

\bibitem{Brenig} M. Tsukada and W. Brenig, Surf. Sci. {\bf 151}, 503 (1985).  

\bibitem{ShinMetiu} S. Shin and H. Metiu, J. Chem. Phys. {\bf 102}, 9285 (1995).

\bibitem{Taneda} A. Taneda, K. Esfarjani, Z.-Q. Li, and Y. Kawazoe, Comp. Mat. Sci. {\bf 9}, 343 (1998).

\bibitem{Boudeville} Y. Boudeville, J. Rousseau-Violet, F. Cyrot-Lackmann, and S. N. Khanna, J. Phys. {\bf 44} 433 (1983).

\bibitem{Pitarke} J. M. Pitarke, V. M. Silkin, E. V. Chulkov, and P. M. Echenique, Rep. Prog. Phys. {\bf 70}, 1 (2007).

\bibitem{Pettifor} D. G. Pettifor, {\it Bonding and Structure of Molecules and Solids} (Oxford University Press, 2002).

\bibitem{surfacebook} See e.g. M. C. Desjonqu\`eres and D. Spanjaard, {\it Concepts in Surface Physics} (Springer, 1996).

\bibitem{Hubbard} J. Hubbard, Proc. Royal Soc. {\bf 276} 238 (1963).

\bibitem{Lang} N.D. Lang, A.R. Williams, Phys. Rev. B {\bf 18}, 616 (1978).

\bibitem{Ueba} H. Ueba, Surf. Sci. Lett. {\bf 215}, 232 (1989).

\bibitem{Almbladh85} C.-O. Almbladh, Phys. Scr. {\bf 32}, 341 (1985).

\bibitem{Pavlyuk15} Y. Pavlyukh, M. Sch\"uler, and J. Berakdar, Phys. Rev. B {\bf 91}, 155116 (2015).

\bibitem{howeverthinfilm} S. Neppl {\it et al.}, Nature {\bf 517}, 342 (2015).

\bibitem{Weightman} P. Weightman, Rep. Prog. Phys. {\bf 45}, 753 (1982). 

\bibitem{CVAuger} C. Verdozzi, A. Marini, and M. Cini, J. Electron. Spectrosc. Relat. Phenom. {\bf 117}, 41 (2005).

\bibitem{Moretti} G. Moretti, Surf. Sci. {\bf 618}, 3 (2013).

\bibitem{BostromPNGF} E. Bostrom, M. Hopjan, A. Kartev, C. Verdozzi, and C.-O. Almbladh, arxiv:1602.07882.

\bibitem{LangFirsov} I.G. Lang, Y. A. Firsov, Zh. Eksp. Teor. Fiz. {\bf 43}, 1843 (1962).

\bibitem{Cini} M. Cini, Phys. Rev.B 17, 2486 (1978).

\bibitem{ParkandLight} T. J. Park and J. C. Light, J. Chem. Phys. {\bf 85}, 5870 (1986).

\bibitem{LiTong86}  T.C. Li and P.Q. Tong, Phys. Rev. A {\bf 34}, 529 (1986).

\bibitem{Tokatly11}  I. V. Tokatly, Phys. Rev. B {\bf 83}, 035127 (2011).

\bibitem{Farzhanepour14} M. Farzanehpour and I. V. Tokatly, Phys. Rev. B {\bf 90}, 195149 (2014).

\bibitem{Bostrompreparation} E. Bostr\"om, A. Mikkelsen, and C. Verdozzi, in progress.

\bibitem{Verdozzi2008} C. Verdozzi, Phys. Rev. Lett. {\bf 101}, 166401 (2008).

\bibitem{Fuks} J. I. Fuks and N. T. Maitra, PCCP {\bf 16}, 14504 (2014).

\bibitem{laser} http://www.laserquantum.com/

\bibitem{Huillier} T. Fordell {\it et al.}, Opt. Express {\bf 17}, 21094 (2009).

\bibitem{Marsell2} E. M\aa rsell {\it et al.}, Nano Lett. {\bf 15} 6601 (2015).

\bibitem{Gadzuk} J. W. Gadzuk, Physica Scripta, {\bf 35}, 171 (1987).

\bibitem{Ivanov} F. Krausz and M. Ivanov, Rev. Mod. Phys. {\bf 81}, 163234 (2009).

\bibitem{Zhou} X. Zhou {\it et al.}, Nature Physics {\bf 8} 232 (2012).

\bibitem{Neutze} R. Neutze {\it et al.}, Nature {\bf 406}, 752 (2000).

\end{thebibliography}
\end{document}